\documentstyle[12pt]{article}

\begin{document}

\begin{flushright}

IMSc/2016/08/03 

\end{flushright} 

\vspace{2mm}

\vspace{2ex}

\begin{center}

{\large \bf A Class of LQC--inspired Models} \\ 

\vspace{4ex}

{\large \bf for Homogeneous, Anisotropic Cosmology} \\ 

\vspace{4ex}

{\large \bf in Higher Dimensional Early Universe } \\

\vspace{8ex}

{\large  S. Kalyana Rama}

\vspace{3ex}

Institute of Mathematical Sciences, HBNI, C. I. T. Campus, 

\vspace{1ex}

Tharamani, CHENNAI 600 113, India. 

\vspace{2ex}

email: krama@imsc.res.in \\ 

\end{center}

\vspace{6ex}

\centerline{ABSTRACT}

\begin{quote} 

The dynamics of a $(3 + 1)$ dimensional homogeneous anisotropic
universe is modified by Loop Quantum Cosmology and,
consequently, it has generically a big bounce in the past
instead of a big-bang singularity. This modified dynamics can be
well described by effective equations of motion. We generalise
these effective equations of motion empirically to $(d + 1)$
dimensions. The generalised equations involve two functions and
may be considered as a class of LQC -- inspired models for $(d +
1)$ dimensional early universe cosmology. As a special case, one
can now obtain a universe which has neither a big bang
singularity nor a big bounce but approaches asymptotically a
`Hagedorn like' phase in the past where its density and volume
remain constant.  In a few special cases, we also obtain
explicit solutions.

\end{quote}

\vspace{2ex}

%Keywords: Loop quantum Cosmology; Higher dimensional cosmology;
%LQC inspired higher dimensional anisotropic cosmology.

\vspace{2ex}

%PACS numbers: 04.50.-h, 04.50.Kd, 04.60.Pp, 11.25.-w

%PACS numbers: 04.50.-h, Higher-dimensional gravity and other
%theories of gravity

%PACS numbers: 04.50.Kd, Modified theories of gravity 

%PACS numbers: 04.60.Pp, Loop quantum gravity, quantum geometry,
%spin foams

%PACS numbers: 11.25.-w, Strings and branes

%PACS numbers: 11.25.Yb, M theory

%PACS numbers: 98.80.Cq Particle-theory and field-theory models
%of the early Universe (including cosmic pancakes, cosmic
%strings, chaotic phenomena, inflationary universe, etc.)

%PACS numbers: 98.80.Qc, Quantum cosmology

\newpage

\vspace{4ex}

\begin{center}

{\bf 1. Introduction} 

\end{center}

\vspace{2ex}

Assume that the universe is $(d + 1)$ dimensional, it is
homogeneous and anisotropic, the $d-$dimensional space is
toroidal, and that the universe is dominated by matter fields,
characterised by density and anisotropic pressures along the $d$
spatial directions. In Einstein's theory, such a universe
generically has a big-bang singularity in the past where its
physical volume vanishes, its density diverges, and the
curvature invariants diverge. In $(3 + 1)$ dimensions, the early
universe dynamics is modified by Loop Quantum Cosmology (LQC)
\cite{b01, abl, aps} which is derived within Ashtekar's Loop
Quantum Gravity (LQG) framework \cite{ashtekar}. See the review
\cite{status} which also contains an exhaustive list of
references. With the dynamics thus modified by quantum effects,
the physical volume of the universe does not vanish but reaches
a non zero minimum, and its density does not diverge but reaches
a finite maximum.  There is no singularity in the past. Instead,
as one goes back in time starting with a large universe, its
volume decreases, reaches a minimum, then bounces back and
increases again.  Correspondingly, the density increases,
reaches a finite maximum, and decreases again. Thus, in LQC, the
$(3 + 1)$ dimensional universe generically has a big bounce in
the past, instead of a big-bang singularity.\footnote{
\label{madh} In a different quantisation scheme \cite{madhavan,
10shyam}, the isotropic evolution of a $(3 + 1)$ dimensional
universe with a massless scalar field has no bounce, but
undergoes a regular non singular extension past the classical
singularity: the volume of the universe vanishes at one time in
the past but its density remains bounded all through the
evolution. However, it is not known whether such an evolution is
generic or whether it is possible in other cases where, for
example, matter content is different or evolution is
anisotropic.} This dynamics, modified by the quantum effects,
can be well described by effective equations of motion which
reduce to the standard Einstein's equations in the `classical
limit'. See \cite{aps, status} for a review.

In this paper, we consider early universe cosmology in $(d + 1)$
dimensional spacetime where $d \ge 3 \;$. We generalise the
effective equations in LQC. These generalised effective
equations of motion may now describe the modified dynamics of a
$(d + 1)$ dimensional homogeneous anisotropic universe. The
generalisation we present involves two functions $f$ and
$\phi^i$, or equivalently $\bar{\mu}^i$, see equations
(\ref{genpsi}) and (\ref{obey}) below. The generalisation is
natural and straightforward but empirical : The functions
involved are not obtained from an underlying theory but are
presented as models. The resulting generalised equations,
involving two functions, may therefore be considered as a class
of LQC -- inspired models for $(d + 1)$ dimensional early
universe cosmology.\footnote{ \label{arnab} Arnab Priya Saha has
pointed out to us that there exists a $(d + 1)$ dimensional LQG
formulation, given in \cite{th1, th2, th3}. Our preliminary
analysis \cite{work} suggests that one can derive the LQC
analogs of the effective equations in $(d + 1)$ dimensions. In
the present empirical framework, this corresponds to a
particular choice of the two functions.}

The matter sector, in both LQC and in the models presented here,
may include various types of scalar fields and other matter
fields. But it is assumed to couple to the metric fields but not
to the curvatures. This assumption also leads to the standard
conservation equation (\ref{t23}) below.

Special cases of the functions in the present models lead to
Einstein's equations in $(d + 1)$ dimensions given, for example,
in \cite{k10} and to the effective LQC equations in $(3 + 1)$
dimensions given, for example, in \cite{status, aw, barrau}. In
the former case, there is a big bang singularity. In the later
case, there is a big bounce. One can also choose the functions
such that the resulting universe has neither a big bang
singularity nor a big bounce but, as one goes back in time
starting with a large volume, the volume approaches a non zero
minimum asymptotically and the density approaches a finite
maximum asymptotically. This is similar to what is expected in
string/M theory where, as one goes back in time, the ten/eleven
dimensional early universe is believed to enter and remain in a
`Hagedorn phase' where its temperature $= {\cal O}(l_s^{- 1})$
and its density $= {\cal O}(l_s^{- (d + 1)}) \;$, $\; l_s$ being
the string length scale \cite{bowick} -- \cite{k07}, \cite{k10}.

The effective equations of motion can be solved numerically for
any choice of the functions. The general features of these
equations, such as when the density will remain finite or will
diverge, can also be seen easily. In general, however, it is not
possible to obtain explicit analytical solutions to the
equations of motion. They can be obtained in some special
cases. In this paper, we will consider a few such cases and
present explicit solutions.

This paper is organised as follows : In Section {\bf 2}, we
present the Einstein's equations for a $(d + 1)$ dimensional
homogeneous anisotropic universe. In Section {\bf 3}, we present
a class of LQC -- inspired models. We give the generalised
effective equations of motion and also show how Einstein's and
effective LQC equations follow as special cases. In Section {\bf
4}, we present a few explicit solutions. In Section {\bf 5}, we
conclude by mentioning a few issues for further studies.

%\newpage

\vspace{4ex}

\begin{center}

{\bf 2. Einstein's equations}

\end{center}

\vspace{2ex}

Let the spacetime be $D = d + 1$ dimensional and let $x^i$, $\;
i = 1, 2, \cdots, d$, denote the spatial coordinates. We take
the $d-$dimensional space to be toroidal and let $L_i$ denote
the coordinate length of the $i^{th}$ direction. Consider the
homogeneous and anisotropic case where the line element $d s$
is given by
\begin{equation}\label{ds}
d s^2 = - d t^2 + \sum_i a_i^2 \; (d x^i)^2
\end{equation}
and the scale factors $a_i \;$ are functions of $t$
only.\footnote{ In the following, the convention of summing over
repeated indices is not always applicable and, hence, will not
be followed. We will write explicitly the indices to be summed
over.} Let the energy momentum tensor $T^A_{\; \; \; B}$, with
$A, B = (0, i) \;$, be diagonal and be given by
\begin{equation}\label{tabdiag}
T^0_{\; \; \; 0} = - \rho \; \; , \; \; \;
T^i_{\; \; \; i} = \hat{p}_i
\end{equation}
where the density $\rho$ is assumed to be positive and the
pressures $\hat{p}_i$ in the $i^{th}$ direction are to be given
by the equations of state. Einstein's equations are given, in
the standard notation with $\kappa^2 = 8 \pi G_D \;$, by
\begin{equation}\label{r11}
R_{A B} - \frac{1}{2} \; g_{A B} R = \kappa^2 \; T_{A B}
\; \; \; , \; \; \; \;
\sum_A \nabla_A T^A_{\; \; \; B} = 0 \; \; .
\end{equation}
Following the notations of our earlier work \cite{k10}, we
define
\[
G_{i j} = 1 - \delta_{i j} \; \; , \; \; \; 
G^{i j} = \frac{1}{d - 1} - \delta^{i j} 
\; \; \; \longleftrightarrow \; \; \; 
\sum_j G^{i j} \; G_{j k} = \delta^i_{\; \; k} 
\]
and
\[
\Lambda = \sum_i \lambda^i \; \; , \; \; \;
\lambda^i_t = \frac{(a_i)_t}{a_i} \; \; , \; \; \;
r^i = \sum_j G^{i j} \; (\rho - \hat{p}_j) = \hat{p}_i
+ \frac{\rho - \sum_j \hat{p}_j}{d - 1} 
\]
where the $t-$subscripts denote time derivatives. Then, after a
straightforward algebra, Einstein's equations (\ref{r11}) give
\footnote{ Note that, written in terms of an average Hubble rate
$\frac{a_t}{a} = \frac {\Lambda_t} {d} \;$ and a shear term
\[
\sigma^2_{shear} = \frac{1}{d \; (d - 1)} \; \sum_{i, j} \left(
\lambda^i_t - \lambda^j_t \right)^2 = \frac {2}{d - 1} \; \left(
\sum_i (\lambda^i_t)^2- \frac{\Lambda_t^2} {d} \right) \; \; , 
\]
equation (\ref{t21}) becomes
\[
\Lambda_t^2 - \sum_i (\lambda^i_t)^2 = d \; (d - 1) \; \left(
\frac{a_t}{a} \right)^2 - \; \left( \frac{d - 1}{2} \right) \;
\sigma^2_{shear} = \; 2 \kappa^2 \; \rho \; \; .
\]
}
\begin{eqnarray}
\sum_{i, j} G_{i j} \; \lambda^i_t \; \lambda^j_t \; = \;
\Lambda_t^2 - \sum_i (\lambda^i_t)^2 & = & 2 \kappa^2 \; \rho
\label{t21} \\
& & \nonumber \\
\lambda^i_{t t} + \Lambda_t \; \lambda^i_t & = & \kappa^2 \; r^i
\label{t22} \\
& & \nonumber \\
\rho_t + \sum_i (\rho + \hat{p}_i) \; \lambda^i_t & = & 0
\; \; . \label{t23}
\end{eqnarray}

Now, define a variable $\tau(t) \;$ by 
\begin{equation}\label{tau}
d t = e^\Lambda \; d \tau \; \; . 
\end{equation}
Then, for any function $X(t)$, equivalently $X(\tau(t))$, we
have
\[
X_\tau = e^\Lambda \; X_t \; \; , \; \; \;
X_{\tau \tau} = 
e^{2 \Lambda} \; \left( X_{t t} + \Lambda_t X_t \right)
\]
where the $\tau-$subscripts denote $\tau$--derivatives. In terms
of $\tau$, equations (\ref{t21}), (\ref{t22}), and (\ref{t23})
become
\begin{eqnarray}
\sum_{i, j} G_{i j} \; \lambda^i_\tau \; \lambda^j_\tau & = &
2 \kappa^2 \; \tilde{\rho} \label{a1} \\
& & \nonumber \\
\lambda^i_{\tau \tau} & = & \kappa^2 \; \tilde{r}^i \; \; .
\label{a2} \\
& & \nonumber \\
\tilde{\rho}_\tau & = & \sum_j (\tilde{\rho} -
\tilde{\hat{p}}_j) \; \lambda^j_\tau \label{a3}
\end{eqnarray}
where
\begin{equation}\label{v2rho}
\tilde{\rho} = e^{2 \Lambda} \; \rho \; \; , \; \; \; 
\tilde{\hat{p}}_i = e^{2 \Lambda} \; \hat{p}_i \; \; , \; \; \;
\tilde{r}^i = e^{2 \Lambda} \; r^i = \sum_j G^{i j} \;
(\tilde{\rho} - \tilde{\hat{p}}_j) \; \; .
\end{equation}

Consider the case where the equations of state are given by
\begin{equation}\label{wi}
\hat{p}_i = w_i \; \rho \; \; \; \longleftrightarrow \; \; \;
\rho - \hat{p}_i = u_i \; \rho
\end{equation}
where $w_i = 1 - u_i$ are constants. Then 
\[
\tilde{r}^i = u^i \; \tilde {\rho} \; \; \; , \; \; \; \; 
u^i = \sum_j G^{i j} \; u_j \; \; . 
\]
Now, in this case, $\lambda^i(\tau)$ and $t(\tau) \;$ can be
obtained explicitly. For this purpose, let
\begin{equation}\label{l}
l = \sum_i u_i \; \lambda^i 
\end{equation}
and let the initial values at an initial time $t_0 \;$ be given
by
\begin{equation}\label{ic}
\left( \lambda^i , \; \lambda^i_t \; ; \; \Lambda , \; l , \;
l_t \; ; \; \rho \; ; \; \tau \right)_{t = t_0} = \; \left(
\lambda^i_0 , \; k^i \; ; \; \Lambda_0 , \; l_0 , \; K \; ; \;
\rho_0 \; ; \; \tau_0 \right)
\end{equation}
where, as follows from the definitions of $\Lambda$ and $l$,
\[
\Lambda_0 = \sum_i \lambda^i_0 \; \; , \; \; \;
l_0 = \sum_i u_i \; \lambda^i_0 \; \; , \; \; \;
K = \sum_i u_i \; k^i 
\]
and $k^i$ and $\rho_0$ must obey equation (\ref{t21}) and,
hence, the relation
\[
\sum_{i, j} G_{i j} k^i k^j = 2 \kappa^2 \; \rho_0 \; \; .
\]
Upon integration, equation (\ref{a3}) gives
\begin{equation}\label{tilderho}
\tilde{\rho} = \tilde{\rho}_0 \; e^{l - l_0} \; \; , \; \; \;
\tilde{\rho}_0 = e^{2 \Lambda_0} \; \rho_0 \; \; . 
\end{equation}
Equations (\ref{a2}) and (\ref{l}) then give an equation for $l
\;$:
\begin{equation}\label{ltautau} 
l_{\tau \tau} = \kappa^2 \; U \; \tilde{\rho}_0 \; e^{l - l_0}
\; \; , \; \; \;
U = \sum_i u_i u^i = \sum_{i, j} G^{i j} u_i u_j
\end{equation}
which, together with the initial values $l_0$ and $K$,
determines $l(\tau) \;$. Equations (\ref{a2}) and (\ref{ic}) now
give
\begin{eqnarray} 
\lambda^i - \lambda^i_0 & = & \frac {u^i} {U} \; (l - l_0) +
e^{\Lambda_0} \; q^i \; (\tau - \tau_0) \label{li} \\
& & \nonumber \\
t - t_0 & = & \int_{\tau_0}^\tau d \tau \; e^\Lambda 
\; \; , \; \; \; \Lambda = \sum_i \lambda^i \label{ttau}
\end{eqnarray}
where
\[
q^i = k^i - \frac{u^i}{U} \; K \; \; \; \longrightarrow \; \; \;
\sum_i u_i q^i = 0 \; \; . 
\]
For stiff matter, the equations of state is given by $\hat{p}^i
= \rho \;$. Hence $u_i = u^i = 0$ and $r^i = 0 \;$. Equations
(\ref{a2}), (\ref{a3}), and (\ref{ic}) then give
\begin{equation}\label{sli} 
\lambda^i - \lambda^i_0 = e^{\Lambda_0} \; k^i \;
(\tau - \tau_0)
\; \; , \; \; \; 
\tilde{\rho} = \tilde{\rho}_0 = e^{2 \Lambda_0} \; \rho_0
\end{equation}
which can also be obtained from equations (\ref{tilderho}) and
(\ref{li}) by setting $u_i = u^i = 0$, thus $l - l_0 = K = 0
\;$. Thus, when the equations of state are linear as in
(\ref{wi}), $\lambda^i(\tau)$ and $t(\tau) \;$ are given
explicitly by equations (\ref{li}), or (\ref{sli}), and
(\ref{ttau}).

%\newpage

\vspace{4ex}

\begin{center}

{\bf 3. A class of LQC--inspired models}

\end{center}

\vspace{2ex}

We first outline briefly the relevant steps involved in
obtaining the effective LQC equations for a $(3 + 1)$
dimensional homogeneous anisotropic universe. We will
then present our generalisations. 

\vspace{4ex}

\begin{center}

{\bf Effective LQC equations} 

\end{center}

\vspace{2ex}

Briefly, the effective equations of motion in LQC may be
obtained as follows. For a detailed derivation and for a
complete description of various terms and concepts mentioned
below, see the review \cite{status}. Let the three dimensional
space be toroidal, the line element $d s$ be given by
\begin{equation}\label{1ds}
d s^2 = - d t^2 + a_1^2 \; (d x^1)^2 + a_2^2 \; (d x^2)^2
+ a_3^2 \; (d x^3)^2
\end{equation}
where $a_i$ are functions of $t$ only, and let $L_i$ and $a_i
L_i$ be the coordinate and the physical lengths of the $i^{th}$
direction. In the LQG formalism, the canonical pairs of phase
space variables consist of an $SU(2)$ connection $A^i_a =
\Gamma^i_a + \gamma K^i_a$ and a triad $E^a_i$ of density weight
one. Here $\Gamma^i_a$ is the spin connection defined by the
triad $e^a_i$, $\; K^i_a$ is related to the extrinsic curvature,
and $\gamma > 0$ and $\approx 0.2375$ is the Barbero -- Immirzi
parameter of LQG, its numerical value being suggested by the
black hole entropy calculations. For the anisotropic universe
whose line element $d s$ is given by equation (\ref{1ds}), one
has
\[
A^i_a \propto c_i \; \; , \; \; \; E^a_i \propto p_i \; \; . 
\]
The full expressions for $A^i_a$ and $E^a_i$ contain various
fiducial triads, cotriads, and other elements. They are given in
\cite{status, aw} but are not relevant for our purposes here
and, hence, not shown. The variables $p_i$ are related to the
scale factors $a_i$ and the lengths $L_i$ by
\[
p_1 = a_2 a_3 \; L_2 L_3 \; \; , \; \; \; 
p_2 = a_1 a_3 \; L_1 L_3 \; \; , \; \; \; 
p_3 = a_1 a_2 \; L_1 L_2
\]
where, with no loss of generality for our purposes here, we have
assumed that the physical and the fiducial triads have same
orientations and hence $p_i$ are all positive. The variables
$c_i \;$ will turn out to be related to $(a_i)_t \;$, see
\cite{aw} and also equation (\ref{cl2}) below. The non vanishing
Poisson brackets among $c_i$ and $p_j$ are given by
\[
\{ c_i, \; p_j \} = \gamma \kappa^2 \; \delta_{i j}
\]
where $\kappa^2 = 8 \pi G_4 \;$. The effective equations of
motion are given by the `Hamiltonian constraint' ${\cal C}_H =
0$ and by the Poisson brackets of $p_i$ and $c_i$ with ${\cal
C}_H$ which give the time evolutions of $c_i$ and $p_i \;$:
namely, by
\[
{\cal C}_H = 0 \; \; , \; \; \;
(p_i)_t = \{ p_i, \; {\cal C}_H \} \; \; , \; \; \;
(c_i)_t = \{ c_i, \; {\cal C}_H \} \; \; . 
\]

Note that it is to be expected that there exists an appropriate
`classical' ${\cal C}_H \;$ the Poisson brackets with which lead
to the classical dynamics. Non trivially, and as reviewed in
detail in \cite{status}, there also exists an effective,
`quantum' modified, ${\cal C}_H$ the Poisson brackets with which
lead to the equations of motion which describe the quantum
dynamics very well. The effective ${\cal C}_H$ reduces to the
classical one in a suitable limit.

The expressions for the classical and the effective ${\cal C}_H
\;$ are given, for example, in \cite{status, aw, barrau}. They
are of the form
\[%begin{equation}\label{1chtot}
{\cal C}_H = H_{grav} (p_i, \; c_i)
+ H_{mat} (p_i \; ; \; \{ \phi_{mat} \}, \; \{ \pi_{mat} \})
\; \; .
\]%end{equation}
The Hamiltonian $H_{mat}$ for a massless scalar field is
considered in \cite{status, aw} and that for a general isotropic
matter is considered in \cite{barrau} \footnote{ The
Hamiltonians ${\cal H}$ and the time variable $\tau$ considered
in these references are related to ours by ${\cal H}
\vert_{theirs} = N H \vert_{ours}$ and $ d t \vert_{ours} = N d
\tau \vert_{theirs} \;$. The lapse function $N = \sqrt{p_1 p_2
p_3}$ for harmonic time $\tau \;$.} where the isotropic pressure
$\hat{P}$ is given by
\begin{equation}\label{pr}
\hat{P} = - \; \frac {\partial H_{mat}}
{\partial \sqrt{p_1 p_2 p_3}} \; \; .
\end{equation}
The classical $H_{grav}$, from which Einstein's equations
follow, is given by
\[
H_{grav} = - \; \frac {c_1 p_1 c_2 p_2 + c_2 p_2 c_3 p_3
+ c_3 p_3 c_1 p_1} {\gamma^2 \kappa^2 \; \sqrt{p_1 p_2 p_3}}
\; \; . 
\]
The effective $H_{grav}$, from which the LQC dynamics follow, is
given in the so--called $\bar{\mu}$ scheme by \footnote{ For an
example of the effective $H_{grav}$ in the so--called $\mu_0$
scheme, see \cite{05shyam}.}
\[
H_{grav} = \frac {- 1} {\gamma^2
\kappa^2 \; \sqrt{p_1 p_2 p_3}} \; \left(
\frac {sin (\bar{\mu}^1 c_1)} {\bar{\mu}^1}  \; 
\frac {sin (\bar{\mu}^2 c_2)} {\bar{\mu}^2}  \; p_1 p_2 
+ \; cyclic \; \; terms \right) 
\]
where 
\[
\bar{\mu}^1 = \lambda_{qm} \; \sqrt{\frac {p_1} {p_2 p_3}}
\; \; , \; \; \; 
\bar{\mu}^2 = \lambda_{qm} \; \sqrt{\frac {p_2} {p_1 p_3}}
\; \; , \; \; \; 
\bar{\mu}^3 = \lambda_{qm} \; \sqrt{\frac {p_3} {p_1 p_2}}
\]
and $\lambda_{qm}^2 = \sqrt{ \frac {3} {4}} \; \gamma \kappa^2$
is the quantum of area. Note that the effective $H_{grav}$
reduces to the classical $H_{grav}$ in the limit $\bar{\mu}^i
c_i \ll 1 \;$ for all $i \;$.

%\newpage

\vspace{4ex}

\begin{center}

{\bf Generalisations} 

\end{center}

\vspace{2ex}

The effective LQC equations for a $(3 + 1)$ dimensional
homogeneous anisotropic universe, given in the previous
subsection, can be generalised straightforwardly to $(d + 1)$
dimensions. We now present these generalised equations. Our
generalisations are empirical and consist of three simple,
straightforward, and natural steps:

\begin{itemize}

\item 

In the LQC expressions given earlier, now let $i = 1, 2, \cdots,
d \;$. The anisotropic pressures $\hat{p}_i$ will be
proportional to the change in energy per unit physical length in
the $i^{th}$ direction. See equation (\ref{phati}) below.

%Also, define the anisotropic pressures so that the classical
%$H_{grav}$ and $H_{mat}$ lead to the equations of motion
%(\ref{t21}) -- (\ref{t23}).

\item

Introduce a function that generalises the trigonometric
functions of the LQC.

\item

Introduce another function that may cover both the $\bar{\mu}$
and the $\mu_0$ schemes.

\end{itemize} 

Our preliminary analysis \cite{work} suggests that, starting
from the $(d + 1)$ dimensional LQG formulation given in
\cite{th1, th2, th3}, it is possible to derive the LQC analogs
of the effective equations in $(d + 1)$ dimensions and, along
the way, the classical equations also. By introducing two
functions, our empirical generalisations here go further beyond
what can be derived within LQC framework. However, it is not
clear to us if they can be obtained from any underlying
theory. In this paper, we simply propose that the resulting
equations may be considered as LQC--inspired models for a $(d +
1)$ dimensional homogeneous anisotropic universe.

We now proceed with the generalisations. Let the $d-$dimensional
space be toroidal. The canonical pairs of phase space variables
are assumed to be given by $c_i$ and $p_i$ where $i = 1, 2,
\cdots, d \;$ now. The variables $p_i \;$, now generalised to
$(d + 1)$ dimensions, are given by \footnote{In the defintion of
$p_i$, we have set the orientation related factors $\epsilon_i =
+ 1$ for all $i \;$ and have assumed that the $p_i$s are all
positive. This suffices for our purposes here.}
\begin{equation}\label{pi}
p_i = \frac {V} {a_i L_i} \; \; , \; \; \;  
V = \prod_j {a_j L_j}
\; \; \; \; \Longrightarrow \; \; \; \; 
V = \left( \prod_i p_i \right)^{\frac {1} {d - 1}} 
\end{equation}
where $L_i$ and $a_i L_i$ are the coordinate and the physical
lengths of the $i^{th}$ direction, and $V$ is the physical
volume. Thus, $p_i$ is the $(d - 1)-$dimensional physical `area'
transverse to the $i^{th}$ direction. Also, define $\lambda^i$
and $l_i$ by
\begin{eqnarray}
& & a_i L_i = e^{\lambda^i} \; \; , \; \; \; V = e^\Lambda
\; \; , \; \; \; p_i = e^{l_i} \nonumber \\
& & \nonumber \\
& \longrightarrow &
l_i = \Lambda - \lambda^i = \sum_j G_{i j} \; \lambda^j
\; \; , \; \; \; \lambda^i = \sum_j G^{i j} \; l_j \; \; .
\label{lambdaidefn} 
\end{eqnarray}

The variables $c_i \;$ will turn out to be related to $(a_i)_t
\;$, see equation (\ref{cl2}) below. We assume that the non
vanishing Poisson brackets among $c_i$ and $p_j$ are given by
\begin{equation}\label{cipi}
\{ c_i, \; p_j \} = \gamma \kappa^2 \; \delta_{i j}
\end{equation}
where $\kappa^2 = 8 \pi G_D \;$ and $\gamma$ is a constant 
parameter.\footnote{ \label{lqm} In the present empirical
framework, the parameter $\gamma$ can be absorbed into the
definition of $c_i \;$. But we retain it explicitly for the sake
of comparisons and also because, in the $(d + 1)$ dimensional
LQG theories \cite{th1, th2, th3}, the parameter $\gamma$
(called $\beta$ in these works) does enter in the definition of
appropriate connection variable. Moreover, in these theories,
$\gamma$ characterises the quantum of the $(d - 1)-$dimensional
`area' which is given by $\lambda_{qm}^{d - 1} = {\cal O}(1) \;
\gamma \kappa^2 \;$. Such a quantisation of area is used in
\cite{nb} to explain the entropy of $(d + 1)$ dimensional black
holes.}  The equations of motion are given by the `Hamiltonian
constraint' ${\cal C}_H = 0 \;$, and by the Poisson brackets of
$p_i$ and $c_i$ with ${\cal C}_H$ which give the evolution of
$c_i$ and $p_i \;$: namely, by
\begin{equation}\label{dynamics}
{\cal C}_H = 0 \; \; , \; \; \;
(p_i)_t = \{ p_i, \; {\cal C}_H \} \; \; , \; \; \;
(c_i)_t = \{ c_i, \; {\cal C}_H \} \; \; . 
\end{equation}
It then follows from equation (\ref{cipi}) that 
\begin{equation}\label{2dynamics}
(p_i)_t = - \; \gamma \kappa^2 \;
\frac {\partial {\cal C}_H} {\partial c_i} \; \; , \; \; \;
(c_i)_t = \gamma \kappa^2 \;
\frac {\partial {\cal C}_H} {\partial p_i} \; \; . 
\end{equation}

Consider ${\cal C}_H \;$. The expressions for ${\cal C}_H \;$
are of the form
\begin{equation}\label{chtot}
{\cal C}_H = H_{grav} (p_i, \; c_i)
+ H_{mat} (p_i \; ; \; \{ \phi_{mat} \}, \; \{ \pi_{mat} \})
\; \; .
\end{equation} 
First consider $H_{mat}$ in equation (\ref{chtot}). It denotes a
generalised matter Hamiltonian which may now include various
types of scalar fields and other matter fields, all symbolically
denoted as $\{ \phi_{mat} \}$ and their conjugate momenta $\{
\pi_{mat} \} \;$. We have assumed that $H_{mat}$ depends only on
$p_i$ and is independent of $c_i \;$. Since $c_i$ will turn out
to be related to $(a_i)_t \;$, this assumption is equivalent to
assuming that matter fields couple to the metric fields but not
to the curvatures. This assumption also leads to the
conservation equation (\ref{t23}) irrespective of what
$H_{grav}$ is : Given $H_{mat}$, define the density $\rho$ and
the pressure $\hat{p}_i$ in the $i^{th}$ direction by
\begin{equation}\label{phati}
\rho = \frac {H_{mat}} {V} \; \; , \; \; \; 
\hat{p}_i = - \; \frac {a_i L_i} {V} \; \frac {\partial H_{mat}}
{\partial (a_i L_i)} = - \; \frac {1} {V} \; \frac {\partial
H_{mat}} {\partial \lambda^i} \; \; .
\end{equation}
The pressure $\hat{p}_i$ is thus, as to be physically expected,
proportional to the change in energy per unit physical length in
the $i^{th}$ direction.\footnote{ If $H_{mat}$ depends on $p_i$
only through volume then its dependence on $\lambda^i$ is only
through $\Lambda = \sum_i \lambda^i \;$. The pressure $\hat{p}_i
= - \; \frac {1} {V} \; \frac {\partial H_{mat}} {\partial
\Lambda} = - \; \frac {\partial H_{mat}} {\partial V} \;$ is
then same for all $i \;$ and is isotropic, {\em cf.} equation
(\ref{pr}).} Differentiating $\rho$, noting that $H_{mat}$ is
independent of $c_i$ and depends only on $p_i$ (equivalently
$\lambda^i$), using $V = e^\Lambda$ and the definition of
$\hat{p}_i$, lead straightforwardly to the conservation equation
(\ref{t23}) : 
\[ 
\rho_t = \left( \frac {H_{mat}} {V} \right)_t 
= - \; \sum_i (\rho + \hat{p}_i) \; \lambda^i_t \; \; .
\]

Next consider $H_{grav}$ in equation (\ref{chtot}). We now
present a straightforward but an empirical generalisation of
those expressions for $H_{grav}$ given in \cite{status, aw,
barrau, 05shyam}, now involving two functions. The resulting
equations of motion will describe a general class of $d + 1$
dimensional homogeneous anisotropic universe. Let $H_{grav}$ be
given by
\begin{equation}\label{hgrav}
H_{grav} = - \; \frac {V \; {\cal G}} {\gamma^2 \lambda_{qm}^2
\kappa^2} \; \; , \; \; \; \; {\cal G} = \frac{1}{2} \sum_{i j}
G_{i j} \psi^i \psi^j = \sum_{i j \; (i < j)} \psi^i \psi^j
\end{equation} 
where $V = \left( \prod_i p_i \right)^{\frac{1}{d - 1}}$, $\;
\lambda_{qm}$ is a length parameter which may be similar to that
mentioned in footnote {\bf \ref{lqm}}, and the fields $\psi^i$
are dimensionless and are to be obtained from an underlying
theory, or to be modelled otherwise. In our model, we assume
that $\psi^i$, $\; i = 1, 2, \cdots, d \;$, are given by
\begin{equation}\label{genpsi} 
\psi^i = \phi^i(p_j) \; f^i \; \; , \; \; \;  \; \; \;
f^i = f(m^i) \; \; , \; \; \;  \; \; \; 
m^i = \bar{\mu}^i(p_j) \; c_i
\end{equation}
where the arguments of the functions $\phi^i, \; f^i$, and
$\bar{\mu}^i$ are as indicated, and 
\begin{equation}\label{obey}
\phi^i \; \bar{\mu}^i = \frac {\lambda_{qm} \; p_i} {V}
\; \; , \; \; \;  \; \; \; 
f(x) \to x \; \; \; as \; \; \;  x \to 0 \; \; . 
\end{equation} 
The conditions (\ref{obey}) on $(\phi^i, \; \bar{\mu}^i, \; f)$
are imposed so that, in the `classical limit' $c_i \to 0 \;$,
one obtains $m^i \to 0 \;$, $\; f^i \to m^i \;$, and
\begin{equation}\label{1einsteinhgrav}
\psi^i = \frac {\lambda_{qm} \; p_i c_i} {V}
\; \; , \; \; \; \; \; 
H_{grav} = - \; \frac {1} {\gamma^2 \kappa^2 V} \;
\sum_{i j \; (i < j)} (p_i c_i) \; (p_j c_j) 
\end{equation}
which, as will be seen below, leads to the Einstein's equations
(\ref{t21}) and (\ref{t22}) for a $(d + 1)$ dimensional
homogeneous anisotropic universe. Thus, the two functions
involved in our generalisation are $f$ and $\phi^i$, or
equivalently $\bar{\mu}^i \;$. For $d = 3 \;$, and $f(x) = sin
\; x \;$, the effective $H_{grav}$ of Loop Quantum Cosmology
given in \cite{status, aw, barrau} follows upon setting $\phi^i
= 1$, and that given in \cite{05shyam} follows upon setting
$\bar{\mu}^i = \epsilon \;$ where $\epsilon$ is a constant,
referred to as the discreteness parameter. 

%\newpage

\vspace{4ex}

\begin{center}

{\bf General Equations of motion}

\end{center}

\vspace{2ex}

Now, from equation (\ref{hgrav}) for $H_{grav}$, it follows that
\begin{eqnarray}\label{dhgravdpc} 
\frac {\partial H_{grav}} {\partial c_i} & = & - \; \frac {V}
{\gamma^2 \lambda_{qm}^2 \kappa^2} \; \frac {\partial {\cal G}}
{\partial c_i} \label{dhgravdc} \\
& & \nonumber \\
- \; \frac {p_i} {V} \; \frac {\partial H_{grav}} {\partial p_i}
& = & \frac {1} {\gamma^2 \lambda_{qm}^2 \kappa^2} \; \left(
\frac {\cal G} { d - 1} + p_i \frac {\partial {\cal G}}
{\partial p_i} \right) \; \; . \label{dhgravdp}
\end{eqnarray}
From $H_{mat}$ being independent of $c_i$, from $\lambda^i =
\sum_j G^{i j} \; (ln \; p_j)$, and from equation (\ref{phati}),
it follows that
\begin{equation}\label{dhmatdpc}
\frac {\partial H_{mat}} {\partial c_i} = 0 \; \; , \; \;  
- \; \frac {p_i} {V} \; \frac {\partial H_{mat}} {\partial p_i}
= \sum_j G^{i j} \hat{p}_j \; \; .
\end{equation}
Then the equations of motion following from equations
(\ref{dynamics}) are given by
\begin{eqnarray}
{\cal G} & = & (\gamma^2 \lambda_{qm}^2 \kappa^2) \; \rho
\label{d1} \\ 
& & \nonumber \\
(ln \; p_i)_t & = &
\left( \frac {V} {\gamma \lambda^2_{qm} p_i} \right) \;
\frac {\partial {\cal G}} {\partial c_i} \label{d2} \\
& & \nonumber 
\end{eqnarray}
\begin{equation}\label{d3}
(ln \; c_i)_t \; = \; \left(
\frac {V} {\gamma \lambda^2_{qm} p_i c_i} \right) \;
\left( \gamma^2 \lambda^2_{qm} \kappa^2 \; R^i - \; p_i
\frac {\partial {\cal G}} {\partial p_i} \right)
\end{equation}
where
\begin{equation}\label{Ri}
R^i = - \; \sum_j G^{i j} \; (\rho + \hat{p}_j) =
r^i - \frac {2 \; \rho} {d - 1} \; \; . 
\end{equation}
Calculating $\rho_t$ using the above equations for $\rho$, $\;
\frac {\partial {\cal G}} {\partial c_i}$ and $\frac {\partial
{\cal G}} {\partial p_i}$, and writing $p_i = e^{l_i} \;$,
gives, as it must, the conservation equation (\ref{t23}) :
\begin{equation}\label{d4}
\rho_t = \sum_i R^i \; (l_i)_t = - \; 
\sum_i (\rho + \hat{p}_i) \; \lambda^i_t \; \; .
\end{equation}

We now specialise to the model where $\psi^i$s are given by
equation (\ref{genpsi}). It is useful to define the following
quantities :
\begin{equation}\label{si}
S_i = \frac {\partial \; {\cal G}} {\partial \psi^i} =
\sum_j G_{i j} \; \psi^j \; \; , \; \; \; \; \; \;
g_i = \frac{d \; f(m^i)} {d m^i} \; \; , \; \; \; \; \; \;
X_i = g_i \; S_i 
\end{equation} 
and 
\begin{equation}\label{nmij}
N^{i j} = \frac{d \; ln \; \phi^i} {d \; ln \; p_j}
\; \; , \; \; \; 
M^{i j} = \frac{d \; ln \; \bar{\mu}^i} {d \; ln \; p_j}
\; \; . 
\end{equation}
From the definitions of $\psi^i$ and ${\cal G}$ and from
equation (\ref{obey}), it follows that 
\begin{eqnarray} 
\sum_i S_i \psi^i = 2 \; {\cal G}
& , &
N^{i j} + M^{i j} + G^{i j} = 0 \nonumber \\
& & \nonumber \\
\frac{d \psi^i} {d c_j} = g_i \left(
\frac {\lambda_{qm} \; p_i} {V} \right) \delta^{i j}
& , & 
\frac{d \; ln \; \psi^i} {d \; ln \; p_j}
= N^{i j} + \frac{ g_i m^i} {f^i} \; M^{i j} \nonumber \\
& & \nonumber \\
\frac {\partial {\cal G}} {\partial c_i} = \left(
\frac {\lambda_{qm} \; p_i} {V} \right) \; X_i
& , & 
p_i \frac {\partial {\cal G}} {\partial p_i} = \sum_j S_j \psi^j
\; \left( N^{j i} + \frac{ g_j m^j} {f^j} \; M^{j i} \right)
\; \; . \label{pdgdp}
\end{eqnarray}
Hence, equation (\ref{d2}) and (\ref{d3}) become
\begin{eqnarray}
(\gamma \lambda_{qm}) \; (ln \; p_i)_t & = & X_i
\; \; \; \longrightarrow \; \; \; \lambda^i_t =
\frac {\sum_j G^{i j} X_j} {\gamma \lambda_{qm}} \label{d2'} \\
& & \nonumber \\
(\gamma \lambda_{qm}) \; (ln \; c_i)_t & = & \frac {1}
{\phi^i m^i} \; \left( \gamma^2 \lambda^2_{qm} \kappa^2 \; R^i
- \; p_i \frac {\partial {\cal G}} {\partial p_i} \right)
\; \; , \label{d3'}
\end{eqnarray}
from which it follows that $\Lambda_t = \frac {\sum_j X_j} {(d -
1) \; \gamma \lambda_{qm}} \;$ and that $(m^i)_t$ is given by 
\begin{eqnarray}
(\gamma \lambda_{qm}) \; \phi^i \; (m^i)_t & = & \gamma^2
\lambda^2_{qm} \kappa^2 \; R^i - \sum_j S_j \psi^j \; N^{j i}
\nonumber \\
& & \nonumber \\
& & \; + \; \sum_j \left( \phi^i m^i M^{i j} - \phi^j m^j
M^{j i} \right) X_j \; \; . \label{d3'mi}
\end{eqnarray}

Thus, equations (\ref{d1}), (\ref{d2'}), and (\ref{d3'}) or
(\ref{d3'mi}) are the independent equations of motion for our
model.\footnote{ Using the chain rule for differentiation, it is
straightforward to obtain the general expressions for $\psi^i_t$
and $(X_i)_t$, and hence for $\lambda^i_t \;$, also. However,
they are not illuminating nor are they useful for our purposes
here and, hence, will not be presented.} They give the time
evolution of all quantities for any given initial values, once
the equations of state are known for the pressures $\hat{p}_i
\;$. Thus, for example, given the values of $(p_i, \; c_i)$ at
some initial time $t_{init} \;$, equation (\ref{d1}) gives
$\rho$, equations of state then give $\hat{p}_i$, equations
(\ref{d2'}) and (\ref{d3'}) give $(p_i)_t$ and $(c_i)_t \;$
which, in turn, give $(p_i, \; c_i)$ at time $t_{init} \pm
\delta t \;$. Repeating these steps gives $(p_i, \; c_i)$ for
all time $t \;$. The $\lambda^i$s follow from equation
(\ref{lambdaidefn}).

Note that if it were possible to invert equation (\ref{d2'}) and
obtain $m^i$ in terms of $\lambda^i \;$, then equations
(\ref{d1}), (\ref{d2'}), and (\ref{d3'mi}) would resemble more
closely the standard FRW equations. This inversion is
generically not possible even for isotropic case and, hence, the
simplest way to understand the evolution equations is as
explained in the previous paragraph.

We will now consider several illustrative cases by making
specific choices for functions $\phi^i, \; \bar{\mu}^i$, and
$f^i = f(m^i) \;$.

%\newpage 

\vspace{4ex}

\begin{center}

{\bf $\mathbf f(m^i) = \alpha \; m^i \;$ and $(d + 1)$
dimensional Einstein's equations}

\end{center}

\vspace{2ex}

Consider first the case where $f^i = f(m^i) = \alpha m^i$ and
$\alpha$ is a constant. Then, for any choice of $\phi^i$ and
$\bar{\mu}^i$ obeying the condition (\ref{obey}), it follows
that $\psi^i \;$, $\; i = 1, 2, \cdots, d \;$, are given by
\begin{equation}\label{falphax}
\psi^i = \alpha \; \frac {\lambda_{qm} \; p_i c_i} {V}
\; \; \; \longrightarrow \; \; \; \; 
H_{grav} = - \; \frac {\alpha^2} {\gamma^2 \kappa^2 V} \;
\sum_{i j \; (i < j)} (p_i c_i) \; (p_j c_j) \; \; .
\end{equation}
Also, $g_i = \alpha$ for all $i \;$. Hence, equation (\ref{d2'})
gives
\begin{equation}\label{cl2}
(\gamma \lambda_{qm}) \; (ln \; p_i)_t = \alpha S_i
\; \; \; \longrightarrow \; \; \;
\lambda^i_t = \frac {\alpha \; \psi^i} {\gamma \lambda_{qm}}
\; \; , \; \; \;
(a_i)_t = \frac {\alpha^2 c_i} {\gamma L_i} \; \; .
\end{equation}
This shows that $c_i$ is related to $(a_i)_t \;$. From equation
(\ref{d3'}), one obtains
\begin{equation}\label{cl3}
(\gamma \lambda_{qm}) \; (ln \; c_i)_t = \frac {\alpha} {\psi^i}
\; \left( \gamma^2 \lambda^2_{qm} \kappa^2 \; r^i
- S_i \; \psi^i \right)
\end{equation}
where, since $g_j m^j = f^j$ now, we have used 
\[
p_i \frac {\partial {\cal G}} {\partial p_i} =
- \sum_j S_j \psi^j \; G^{i j} =
- \; \frac {2 \; {\cal G}} {d - 1} + S_i \psi^i \; \; .
\]
One can now calculate $(ln \; \psi^i)_t$ and, after a little
algebra, it follows that
\begin{equation}\label{clpsi}
\psi^i_t + \Lambda_t \; \psi^i =
\alpha \; (\gamma \lambda_{qm} \kappa^2) \; r^i \; \; .
\end{equation}
Substituting $\alpha \psi^i = (\gamma \lambda_{qm}) \;
\lambda^i_t \;$, equations (\ref{d1}) and (\ref{clpsi}) give
\begin{eqnarray}
\sum_{i, j} G_{i j} \; \lambda^i_t \; \lambda^j_t & = & 
2 \alpha^2 \kappa^2 \; \rho \label{alpha1} \\
& & \nonumber \\
\lambda^i_{t t} + \Lambda_t \; \lambda^i_t & = &
\alpha^2 \kappa^2 \; r^i \label{alpha2} \; \; . 
\end{eqnarray}
For $\alpha = 1$, these are indeed the Einstein's equations
(\ref{t21}) and (\ref{t22}) for a $(d + 1)$ dimensional
homogeneous anisotropic universe. These equations may also be
thought of as Einstein's equations with $t$ replaced by $\alpha
t \;$. Thus, for example, $\alpha = - 1$ can be thought of as
reversing the direction of time.

%\newpage 

\vspace{4ex}

\begin{center}

{\bf $\mathbf f(m^i) = \alpha \; (m^i  - m_{shift}) \;$}

\end{center}

\vspace{2ex}

Consider now the case where the function
\[
f^i = f(m^i) = \alpha \; \tilde{m}^i
\; \; \; , \; \; \; \; \; \; \tilde{m}^i = m^i - m_{shift}
\]
and $\alpha$ and $m_{shift}$ are constants, same for all $i
\;$. The equations (\ref{d1}) and (\ref{d2'}) will remain the
same when expressed in terms of $p_i$ and $\tilde{m}^i \;$.
However, equation (\ref{d3'mi}) for $(\tilde{m}^i)_t \;$ will,
in general, be different and will have explicit dependence on
the shift constant $m_{shift} \;$. In the two examples of
$\phi^i$ to be given below, this dependence will drop out. In
these examples then, but not in general, $f(m^i) = \alpha \;
\tilde{m}^i$ will again lead to the Einstein's equations for a
$(d + 1)$ dimensional homogeneous anisotropic universe.

%\newpage

\vspace{4ex}

\begin{center}

{\bf LQC--inspired equations for $(d + 1)$ dimensional
cosmology}

\end{center}

\vspace{2ex}

%\newpage

Although introducing the function $\phi^i$ renders our model in
equation (\ref{genpsi}) more general, it does not seem to be of
much help in obtaining analytical solutions. Therefore, we will
consider only two explicit examples of $\phi^i \;$ : One, as in
\cite{status, aw, barrau},
\begin{equation}\label{ex1}
{\mathbf (1)}  \; \; \;  \; \; \;
\phi^i = 1 \; \; , \; \; \;  \; \; \;
\bar{\mu}^i = \frac {\lambda_{qm} \; p_i} {V}
\; \; \; \;  \longrightarrow \; \; \; \;
N^{i j} = 0 
\end{equation}
and another, as in \cite{05shyam}, 
\begin{equation}\label{ex2}
{\mathbf (2)}  \; \; \;  \; \; \;
\phi^i = \frac {\lambda_{qm} \; p_i} {\epsilon \; V} 
\; \; , \; \; \;  \; \; \;
\bar{\mu}^i = \epsilon 
\; \; \; \;  \longrightarrow \; \; \; \;
M^{i j} = 0 
\end{equation}
where $\epsilon$ is a constant, referred to as the discreteness
parameter. Note that in these two examples, the equations remain
the same under a shift of $m^i$ where $m^i \to \tilde{m}^i = m^i
- m_{shift}$ and $m_{shift}$ is a constant, same for all $i \;$.
This is because in example {\bf (1)}, $\; \phi^i = 1$ and $M^{i
j} = M^{j i} = - G^{i j} \;$, and the right hand side of
equation (\ref{d3'mi}) depends only on the difference $(m^i -
m^j) = (\tilde{m}^i - \tilde{m}^j) \;$. In example {\bf (2)},
$M^{i j} = 0$ and the right hand side of equation (\ref{d3'mi})
has no explicit dependence on $m^i$s. We now consider these two
examples in more detail.

%\newpage

\vspace{4ex}

\begin{center}

{\bf Example $\mathbf (1) : \; \; \; \; \phi^i = 1
\; \; \; , \; \; N^{i j} = 0$}

\end{center}

\vspace{2ex}

In this case where $\phi^i = 1$, we have $\psi^i = f^i \;$ and
$\bar{\mu}^i = \frac {\lambda_{qm} \; p_i} {V} \;$. This
expression for $\bar{\mu}^i$ is a $(d + 1)$ dimensional
generalisation of that given in \cite{status, aw, barrau}.
Using $\phi^i = 1 \;$, $\; M^{i j} = - \; G^{i j} \;$, and
\[
\sum_j G^{i j} \; (m^i - m^j) \; X_j \; = \; 
\frac {\sum_j (m^i - m^j) \; X_j} {d - 1}
\]
in equation (\ref{d3'mi}), it follows that 
\begin{equation}\label{d3(1)}
(m^i)_t + \frac {\sum_j (m^i - m^j) \; X_j} {(d - 1) \;
\gamma \lambda_{qm}} \; = \; \gamma \lambda_{qm} \kappa^2 \; R^i
\; \; .
\end{equation}
Thus, equations (\ref{d1}), (\ref{d2'}), and (\ref{d3(1)}) may
be taken to be the equations of motion in the case where $\phi^i
= 1 \;$.

Note that setting $f^i = sin \; m^i$, $\; i = 1, 2, \cdots, d
\;$, gives the analog of the effective LQC equations,
generalised now to a $(d + 1)$ dimensional homogeneous
anisotropic universe. The corresponding $H_{grav}$ is given by
\footnote{Our preliminary analysis \cite{work} suggests that
these $(d + 1)$ dimensional LQC analogs with $f^i = sin \; m^i$
can be derived from an underlying theory, namely from the $(d +
1)$ dimensional LQG formulation given in \cite{th1, th2,
th3}. Isotropic case arising from this formulation has been
considered in \cite{zhang}.}
\[
H_{grav} = - \; \frac {V} {\gamma^2 \lambda_{qm}^2 \kappa^2} \;
\sum_{i j \; (i < j)} (sin \; m^i) \; (sin \; m^j) \; \; .
\]

Also note that, when $d = 3$ and $f^i = sin \; m^i \;$, the
expression for $(c_i)_t$ agrees with that given in \cite{aw} for
$\hat{p}_i = \rho \;$ which is the equation of state for a
massless scalar field considered there; and, that the expression
for $(m^i)_t$ agrees with that given in \cite{barrau} for
isotropic pressures, namely for $\hat{p}_i = \hat{p} \;$.

%\newpage

\vspace{4ex}

\begin{center}

{\bf Example $\mathbf (2) : \; \; \; \; \bar{\mu}^i =
\epsilon \; \; \; , \; \; M^{i j} = 0$}

\end{center}

\vspace{2ex}

In this case we have $\phi^i = \frac {\lambda_{qm} \; p_i}
{\epsilon \; V}$ and $\psi^i = \phi^i f^i \;$ where $\epsilon$
is a constant, referred to as the discreteness parameter in
\cite{05shyam}. Using $N^{i j} = - \; G^{i j} \;$,
\[
\sum_j G^{i j} (S_j \psi^j) \; = \;
\frac {2 \; {\cal G}} {d - 1} - S_i \psi^i  \; \; , 
\]
and $R^i + \frac {2 \rho} {d - 1} = r^i \;$ in equation
(\ref{d3'mi}), it follows that
\begin{equation}\label{dmidt2}
\gamma \lambda_{qm} \; \phi^i \; (m^i)_t \; = \; (\gamma^2
\lambda^2_{qm} \kappa^2) \; r^i - S_i \psi^i 
\end{equation}
and, after a little algebra, further that
\begin{equation}\label{d3(2)}
\psi^i_t + \Lambda_t \; \psi^i \; = \;
(\gamma \lambda_{qm} \kappa^2) \; g_i \; r^i \; \; .
\end{equation}
Thus, equations (\ref{d1}), (\ref{d2'}), and (\ref{d3(2)}) may
be taken to be the equations of motion in the case where
$\bar{\mu}^i = \epsilon \;$.

Note that, when $d = 3$ and $f^i = sin \; m^i \;$, these
expressions agree with those given in \cite{05shyam} for
$\hat{p}_i = \rho = 0 \;$.

\vspace{4ex}

\begin{center}

{\bf Isotropic case}

\end{center}

\vspace{2ex}

Consider the isotropic case where 
\[
\hat{p}^i = \hat{p} \; \; , \; \; \;
(p^i, \; c_i) = (p, \; c)
\; \; , \; \; \; a_i = a \; \; . 
\]
Then for any choice of $\phi^i = \phi$ and $\bar{\mu}^i =
\bar{\mu}$, we have, for example,
\[
(m^i; \; \psi^i; \; g_i, \; S_i, \; X_i) \; = \;
(m; \; \psi; \; g, \; S, \; X) \; \; , \; \; \;
\; \; \;  \; \; \;
\lambda^i_t \; = \; \frac{a_t}{a} 
\]
and, hence, 
\[
2 \; {\cal G} = d (d - 1) \; \psi^2 
\; \; , \; \; \; 
S = (d - 1) \; \psi
\; \; , \; \; \; 
X = g \; S \; \; . 
\]
Equations (\ref{d1}) and (\ref{d2'}) then give
\begin{equation}\label{iso1}
\psi^2 \; = \; \frac {2 \; \gamma^2 \lambda_{qm}^2 \kappa^2}
{d \; (d - 1)} \; \rho \; = \; \frac {\rho} {\rho_{ub}} 
\; \; , \; \; \;  \; \; \;
\frac{a_t}{a} \; = \; \frac {g \; \psi} {\gamma \lambda_{qm}}
\end{equation}
where $\rho_{ub} = \frac {d \; (d - 1)} {2 \; \gamma^2
\lambda_{qm}^2 \kappa^2} \;$. One now obtains
\begin{equation}\label{iso12}
\left( \frac{a_t}{a} \right)^2 \; = \;
\frac {2 \; \kappa^2} {d \; (d - 1)} \; (\rho \; g^2) \; \; .
\end{equation}
Equation (\ref{d3'}) or (\ref{d3'mi}) gives an expression for
$c_t \;$ or $m_t \;$. To proceed further, $\phi$ and $\bar{\mu}$
are needed. In example {\bf (1)} where $\phi = 1$, it follows
from equation (\ref{d3(1)}) that
\begin{equation}\label{d3iso}
m_t \; = - \; \frac {\gamma \lambda_{qm} \kappa^2} {d - 1} \;
(\rho + \hat{p})
\end{equation}
where we have used $m^i = m$ and $R^i = R = - \; \frac {\rho +
\hat{p}} {d - 1} \;$. Also, note that when $\phi = 1$ and
$\psi(m) = f(m) = sin \; m \;$, we have $g^2 = cos^2 \; m = 1 -
\frac {\rho} {\rho_{ub}} \;$. Equation (\ref{iso12}) then gives
the $(d + 1)$ dimensional result obtained in \cite{zhang}.

\newpage

\vspace{4ex}

\begin{center}

{\bf 4. A few explicit solutions}

\end{center}

\vspace{2ex}

Equations of motion (\ref{d1}), (\ref{d2'}), and (\ref{d3'}) can
be solved numerically for any choice of the functions $\phi^i$
and $f(x) \;$, as explained below equation (\ref{d3'mi}). The
general features of these equations can also be seen easily.
For example: {\bf (a)} In the limit $f(x) \to x \;$, the
equations of motion become same as Einstein's equations. {\bf
(b)} The density $\rho$ is bounded from above if the functions
$\psi^i$ are. That is,
\[
\psi^i \le \psi_{mx}  \; \; \; \Longrightarrow \; \; 
\rho \le \rho_{ub} 
\]
where $\rho_{ub} = \frac {d \; (d - 1)} {2 \; \gamma^2
\lambda_{qm}^2 \kappa^2} \; (\psi^2_{mx}) \;$. In example {\bf
(1)}, $\psi^i = f^i = sin \; m^i$ and, hence, $\psi_{mx} = 1
\;$. In a given anisotropic evolution $\rho$ may not reach
$\rho_{ub}$ since not all $\psi^i$ may reach the maximum value
$\psi_{mx}$ at the same time.

In general, it is not possible to obtain explicit analytical
solutions to the equations of motion. They can, however, be
obtained in a few special cases. We will now consider such
cases.

%\newpage

\vspace{2ex}

\begin{center}

{\bf Anisotropic case in Example (2) with stiff matter} 

\end{center}

\vspace{2ex}

Consider the anisotropic case in example {\bf (2)} where
$\bar{\mu}^i = \epsilon \;$. Let the equations of state be given
by that of a stiff matter, namely by $\hat{p}^i = \rho \;$.
\footnote{ There is not much loss of generality in considering
stiff matter equations of state. This is because the
modfications in the effective equations are expected to become
important only when the volume is small and the densities are
high. In this limit, the dominant matter fields will be those
with the highest $w = \frac {\hat{p}^i} {\rho} \;$, namely $w =
1 \;$.} Then $r^i = 0 \;$ and equation (\ref{d3(2)}) can be
integrated to obtain
\begin{equation}\label{21}
\psi^i = \frac {\lambda_{qm}} {\epsilon \; V} \; K^i
\; \; \; \longleftrightarrow \; \; \; 
f^i = f(m^i) = \frac {K^i} {p_i} \; \; . 
\end{equation}
where $K^i$ are integration constants. With no loss of
generality, we assume that $K^i$ are all strictly positive,
namely that $K^i > 0$ for all $i \;$. Note then that $p_i$s are
bounded from below if the functions $f^i$ are bounded from
above. That is,
\[
f^i \le f_{mx}  \; \; \; \Longrightarrow \; \; \; 
p_i \ge \frac {K^i} {f_{mx}} \; \; . 
\]
Also, we have
\begin{equation}\label{22}
S_i = \frac {\lambda_{qm}} {\epsilon \; V} \; \sigma^i
\; \; , \; \; \; \; \sigma_i = \sum_j G_{i j} \; K^j
\; \; , \; \; \; \; X_i = g_i \; S_i \; \; .
\end{equation}
Equation (\ref{d2'}) then gives
\begin{equation}\label{23}
\frac {d p_i} {g_i \; p_i} = 
\frac {\sigma^i} {\gamma \; \epsilon} \; d \tau
\; \; , \; \; \; \;
d t = V \; d \tau \; \; .
\end{equation}

Thus, for a given function $f(x)$, inverting equation (\ref{21})
gives $m^i$ and then $g_i = \frac {d f(m^i)} {d m^i} \;$ in
terms of $K^i$ and $p_i \;$. Then integrating equation
(\ref{23}) gives $p_i$ in terms of $\tau \;$. Using $V = \left(
\prod_i p_i \right)^{\frac{1}{d - 1}} \;$ in equation (\ref{23})
gives $t$ in terms of $\tau \;$. Then equation (\ref{pi}) gives
the scale factors $a_i \;$ and equation (\ref{d1}) gives $\rho
\;$ :
\begin{equation}\label{airho}
a_i = \frac {V} {p_i L_i} \; \; , \; \; \;
\rho = \left( \frac {\sum_{i j} G_{i j} K^i K^j}
{2 \; \gamma^2 \epsilon^2 \kappa^2} \right) \; \frac {1} {V^2}
\; \; . 
\end{equation}
It follows from the above expressions that if the $p_i$s are
bounded from below by $p_{min} > 0$, that is if $p_i \ge p_{min}
> 0$, then the volume $V$ will not vanish and the density $\rho$
will remain finite and not diverge.

%\newpage

\vspace{2ex}

\begin{center}

{\bf (i) $\; \; \mathbf  f(m^i) = sin \; m^i \;$}

\end{center}

\vspace{2ex}

Consider the case where $f^i = f(m^i) = sin \; m^i \;$. In this
case, we have
\[
g_i = cos \; m^i \; \; , \; \; \;
g_i \; p_i = \sqrt{p_i^2 - (K^i)^2} \; \; . 
\]
Equation (\ref{23}) for $p_i$ can be easily integrated and one
obtains
\[
p_i = K^i \; cosh \; \theta_i \; \; , \; \; \; 
sin \; m^i = \frac {1} {cosh \; \theta_i} \; \; . 
\]
where $\theta_i(\tau)$ are given by
\[
\theta_i = \frac {\sigma^i} {\gamma \; \epsilon} \;
(\tau - \tau_0) + \theta_{i 0} = \frac {\sigma^i}
{\gamma \; \epsilon} \; (\tau - \tau_i) \; \; .
\]
In the above equation, we have assumed that $p_i = p_{i 0} = K^i
\; cosh \; \theta_{i 0}$ at an initial time $\tau_0 \;$, and the
second equality defines $\tau_i \;$. Let $t = t_0$ be the
initial time at $\tau_0 \;$. Then equation (\ref{23}) for $t$
gives
\[
t - t_0 = \int^\tau _{\tau_0} \; d \tau \;
\left( \prod_i p_i \right)^{\frac {1} {d - 1}} \; \; .
\]

Note that $ p_i(\tau) \ge p_i(\tau_i) = K^i > 0 \;$. Hence it
follows that the volume $V$ will not vanish and, from equation
(\ref{airho}), that the density $\rho$ will not diverge. It is
straightforward to see that the variable $\tau$ can range
between $- \infty$ and $+ \infty \;$; the $\theta_i$s and $t$
range between $- \infty$ and $+ \infty \;$; and the $m^i$s
between $0 \;$ and $\pi \;$. When all the $m^i$s are near $0$ or
$\pi \;$, we have $g_i = + 1$ or $ - 1$ for all $i \;$. The
evolution is then same as that given by Einstein's equations.
The precise details of the evolution depend on the initial
values $K^i \;$.

%\newpage

\vspace{2ex}

\begin{center}

{\bf (ii) $\; \; \mathbf f(m^i) = \frac {\alpha m^i} {m_*} \;
\left( 2 - \frac {m^i} {m_*} \right) \;$}

\end{center}

\vspace{2ex}

A closer inspection of the results in the previous case shows
that the salient features of the evolution there are due to the
fact that the function $f(x) = sin \; x$ starts lineary near $x
= 0 \;$; reaches a maximum; and then decreases and reaches a
zero again linearly at $x = \pi \;$. This suggests that any
function with these properties must also result in similar
salient features of the evolution. In order to illustrate this
explicitly, consider a case with these properties where now $f^i
= f(m^i) = \frac {\alpha m^i} {m_*} \; \left( 2 - \frac {m^i}
{m_*} \right) \;$ and $\; \alpha$ and $m_*$ are positive
constants, same for all $i \;$. In this case, we have
\[
g_i = \frac {2 \alpha} {m_*} \; \left( 1 - \frac {m^i} {m_*}
\right) = \frac {2 \alpha} {m_*} \; \epsilon_i \; 
\sqrt{1 - \frac {f^i} {\alpha}}
\]
where $\epsilon_i = sgn \left( 1 - \frac {m^i} {m_*} \right) \;$
and
\[
g_i \; p_i = \frac {2 \alpha} {m_*} \; \epsilon_i \; 
\sqrt{p_i^2 - \frac {K^i p_i} {\alpha}} \; \; .
\]
Equation (\ref{23}) for $p_i$ can be easily integrated and,
after a little algebra, one obtains
\[
p_i = \frac {K^i} {\alpha} \; cosh^2 \; \theta_i
\]
where $\theta_i(\tau)$ are given by
\[
\theta_i = \frac {\alpha} {m_*} \; \frac {\sigma^i}
{\gamma \; \epsilon} \; (\tau - \tau_0) + \theta_{i 0} =
\frac {\alpha} {m_*} \; \frac {\sigma^i} {\gamma \; \epsilon}
\; (\tau - \tau_i) \; \; .
\]
In the above equation, we have assumed that $p_i = p_{i 0} =
\frac {K^i} {\alpha} \; cosh^2 \; \theta_{i 0}$ at an initial
time $\tau_0 \;$, and the second equality defines $\tau_i
\;$. Let $t = t_0$ be the initial time at $\tau_0 \;$. Then
equation (\ref{23}) gives
\[
t - t_0 = \int^\tau _{\tau_0} \; d \tau \;
\left( \prod_i p_i \right)^{\frac {1} {d - 1}} \; \; .
\]

Note that $ p_i(\tau) \ge p_i(\tau_i) = \frac {K^i} {\alpha} > 0
\;$. Hence it follows that the volume $V$ will not vanish and,
from equation (\ref{airho}), that the density $\rho$ will not
diverge. It is straightforward to see that the variable $\tau$
can range between $- \infty$ and $+ \infty \;$; the $\theta_i$s
and $t$ range between $- \infty$ and $+ \infty \;$; and the
$m^i$s between $0 \;$ and $2 m_* \;$. When all the $m^i$s are
near $0$ or $2 m_* \;$, we have $g_i = + \frac {2 \alpha} {m_*}$
or $- \frac {2 \alpha} {m_*}$ for all $i \;$. The evolution is
then same as that given by Einstein's equations. The precise
details of the evolution depend on the initial values $K^i \;$.

%\newpage

\vspace{2ex}

\begin{center}

{\bf (iii) $\; \; \mathbf f(m^i) = \alpha \; \left( 1 -
\left( 1 - \frac {m^i} {m_*} \right)^{2 n} \right) \;$}

\end{center}

\vspace{2ex}

We now consider another case where $f(m^i)$ starts from zero,
reaches a maximum, and falls back to zero, but now its shape
near the maximum can be made flatter. Hence, consider a case
where now $f^i = f(m^i) = \alpha \; \left( 1 - \left( 1 - \frac
{m^i} {m_*} \right)^{2 n} \right) \;$, $\; \alpha$ and $m_*$ are
positive constants, same for all $i \;$, and $\; n$ is a
positive integer. The previous example corresponds to $n = 1
\;$. The maximum, which is at $m_* \;$, will be flatter for
larger values of $n \;$. In this case, we have

\[
g_i = \frac {2 n \alpha} {m_*} \; \left( 1 - \frac {m^i} {m_*}
\right)^{2 n - 1} = \frac {2 n \alpha} {m_*} \; \epsilon_i \; 
\left( 1 - \frac {f^i} {\alpha} \right)^{\frac {2 n - 1} {2 n}}
\]
where $\epsilon_i = sgn \left( 1 - \frac {m^i} {m_*} \right) \;$
and
\[
g_i \; p_i = \frac {2 n \alpha} {m_*} \; \; \epsilon_i \;
p_i^{\frac {1} {2 n}} \; \left( p_i - \frac {K^i} {\alpha}
\right)^{\frac {2 n - 1} {2 n}} \; \; .
\]

We are not able to integrate equation (\ref{23}) for $p_i$ now
for arbitrary values of integer $n \;$. However, one can obtain
straightforwardly the leading behaviour near the zeros and the
maximum of $f^i \;$, namely near $m^i = 0, \; m_*$, and $2 m_*
\;$. Near $m^i = 0 \;$, the function $f$ is of the type $f(m^i)
= \alpha m^i \;$ and, hence, the earlier analysis carries
over. Near $m^i = 2 m_* \;$, the function $f$ is of the type
$f(m^i) = \alpha (m^i - m_{shift}) \;$. Note that for both the
anisotropic examples {\bf (1)} and {\bf (2)}, the form of the
equations of motion, see equations (\ref{d3(1)}) and
(\ref{dmidt2}) for $(m^i)_t$ in particular, remain the same
under a constant shift in $m^i$ with the shift being the same
for all $i \;$. It is then clear that, after a constant shift in
$m^i \;$, the earlier analysis carries over near $m^i = 2 m_*
\;$ also.

Consider when $m^i$, for a given $i \;$, is near $m_* \;$ and
$\tau$ is near $\tau_i \;$ where $\tau_i$ is defined by
$m^i(\tau_i) = m_* \;$. The corresponding $f^i \stackrel {<}
{_\sim} \alpha \;$ and, in the limit $\tau \simeq \tau_i \;$,
let
\[
p_i = \frac {K^i} {\alpha} \; (1 + x^i)
\; \; \; , \; \; \; \; \; x^i \ll 1 \; \; .
\]
One can then integrate equation (\ref{23}) in this limit. After
a straightforward algebra, it follows that
\[
x^i = \left( \frac {\alpha \; \sigma^i} {m_* \; \gamma \;
\epsilon} \right)^{2 n} \; (\tau - \tau_i)^{2 n} \; \; .
\]

Note that $ p_i(\tau) \ge p_i(\tau_i) = \frac {K^i} {\alpha} > 0
\;$. Hence it follows that the volume $V$ will not vanish and,
from equation (\ref{airho}), that the density $\rho$ will not
diverge. It can be seen by a straightforward but qualitative
analysis that the variable $\tau$ can range between $- \infty$
and $+ \infty$, $\; t$ ranges between $- \infty$ and $+ \infty
\;$, and the $m^i$s between $0 \;$ and $2 m_* \;$. When all the
$m^i$s are near $0$ or $2 m_* \;$, we have $g_i = + \frac {2 n
\alpha} {m_*}$ or $- \frac {2 n \alpha} {m_*}$ for all $i \;$.
The evolution is then same as that given by Einstein's
equations. The precise details of the evolution depend on the
value of $n$ and the initial values $K^i \;$.

%\newpage

\vspace{2ex}

\begin{center}

{\bf (iv) $\; \; \mathbf f(m^i) = \alpha \; (1 - e^{- \beta
m^i}) \;$}

\end{center}

\vspace{2ex}

We now consider another case where $f(m^i)$ starts from zero and
increases monotonically to a constant value. Hence, consider a
case where now $f^i = f(m^i) = \alpha \; (1 - e^{- \beta m^i})
\;$ and $\; \alpha$ and $\beta$ are positive constants, same for
all $i \;$. In this case, we have
\[
g_i = \alpha \beta \; e^{- \beta m^i} = \beta \; (\alpha - f^i)
\; \; , \; \; \;
g_i \; p_i = \beta \; (\alpha p_i - K^i) \; \; . 
\]
Equation (\ref{23}) for $p_i$ can be easily integrated and,
incorporating the condition that $p_i = p_{i 0}$ at an initial
time $\tau_0 \;$, one obtains
\[
ln \; (\alpha p_i - K^i) = \frac {\alpha \beta \; \sigma^i}
{\gamma \; \epsilon} \; (\tau - \tau_0)
+ ln \; (\alpha p_{i 0} - K^i) \; \; . 
\]
Let $t = t_0$ be the initial time at $\tau_0 \;$. Then equation
(\ref{23}) gives
\[
t - t_0 = \int^\tau _{\tau_0} \; d \tau \;
\left( \prod_i p_i \right)^{\frac {1} {d - 1}} \; \; .
\]
Note that $ p_i(\tau) = \frac {K^i} {f^i} \ge p_i(- \infty) =
\frac {K^i} {\alpha} > 0 \;$. Hence it follows that the volume
$V$ will not vanish and, from equation (\ref{airho}), that the
density $\rho$ will not diverge. Also, it is straightforward to
see that as $p_i$ ranges between $+ \infty$ and $\frac {K^i}
{\alpha} \;$, the variables $\tau$ and $t$ range between $+
\infty$ and $- \infty$, and the $m^i$s range between $0 \;$ and
$+ \infty \;$. When all the $m^i$s are near $0$ or $+ \infty
\;$, we have $g_i = \alpha \beta $ or $0$ for all $i \;$. The
evolution is then same as that given by Einstein's equations.
The precise details of the evolution depend on the initial
values $K^i \;$. However, note that, as $m^i \to \infty$,
$f(m^i)$ in this case approaches the maximum asymptotically and
does not decrease from its maximum. As $\tau \to - \infty$, it
is easy to see that all the $m^i$s approach $\infty$ and $g_i$s
all vanish and, hence, that the density $\rho$ and the scale
factors $a^i$ reach their finite, non zero constant values
asymptotically. Consequently, there is no bounce where the scale
factors increase again from their minimum. This phase of the
evolution, where the density and volume remain constant, is
similar to the `Hagedorn phase' in string/M theory where, as one
goes back in time, the universe's temperature $\to {\cal
O}(l_s^{- 1})$ and its density $\to {\cal O}(l_s^{- (d + 1)})
\;$ asymptotically, $\; l_s$ being the string length scale
\cite{bowick} -- \cite{k07}, \cite{k10}.

%\newpage

\vspace{2ex}

\begin{center}

{\bf Isotropic case in Example (1)} 

\end{center}

\vspace{2ex}

Consider the isotropic case in example {\bf (1)} where $\phi =
1$ and $\psi = f(m) \;$. Let the equation of state be given by
$\hat{p} = w \; \rho$ where $w$ is a constant. Then equations
(\ref{iso1}) and (\ref{d3iso}) give
\begin{equation}\label{1iso1}
- \; \frac {d m} {f^2} = \frac {d \; (1 + w)}
{2 \; \gamma \lambda_{qm}} \; d t \; \; , \; \; \; \;
\frac{d a}{a} = \frac {g \; f} {\gamma \lambda_{qm}} \; d t
\; \; .
\end{equation}
Thus, for a given function $f(m)$, integrating equation
(\ref{1iso1}) gives $m$ in terms of $t \;$. Defining ${\cal
F}(m)$ and incorporating the conditions that $m = m_0$ and
${\cal F} = {\cal F}_0$ at an initial time $t = t_0 \;$, we
write
\begin{equation}\label{1iso2}
{\cal F} = - \; \int \frac {d m} {f^2} \; = \; 
\frac {d \; (1 + w)} {2 \; \gamma \lambda_{qm}} \;
(t - t_0) + {\cal F}_0 \; \; . 
\end{equation}
Sometimes, ${\cal F}$ may also be written as 
\[
{\cal F} = \frac {d \; (1 + w)} {2 \; \gamma \lambda_{qm}} \;
(t - t_{ub})
\]
where $t = t_{ub}$ is the time when ${\cal F} = 0 \;$. One can
now obtain $f(m)$ and $g(m) = \frac {d f} {d m}$ in terms of
$t \;$. Then another integration will give the scale factor
$a(t) \;$. Or, alternatively, equations (\ref{d4}) and
(\ref{iso1}) give the scale factor $a$ directly in terms of $f
\;$:
\begin{equation}\label{1iso3}
\rho \; = \; \rho_0 \; \left( \frac {a} {a_0}
\right)^{- d (1 + w)} \; = \; \rho_{ub} \; f^2
\end{equation}
where $\rho_{ub} = \frac {d \; (d - 1)} {2 \; \gamma^2
\lambda_{qm}^2 \kappa^2} \;$ and we have incorporated the
condition that $\rho = \rho_0$ and $a = a_0$ at an initial time
$t = t_0 \;$. Thus, for a given $f(m)$, finding ${\cal F}$ and
expressing $f$ in terms of ${\cal F}$, where possible,
constitute an explicit solution.

%\newpage

\vspace{2ex}

\begin{center}

{\bf (i) $\; \; \mathbf  f(m) = sin \; m \;$}

\end{center}

\vspace{2ex}

In the case where $f(m) = sin \; m \;$, one obtains that
\[
{\cal F} = cot \; m 
\; \; \; , \; \; \;  \; \; \; 
f^2 = \frac {1} {1 + {\cal F}^2} \; \; .
\]
Note that ${\cal F}(t)$ and $f(t)$ are of the form
\[
{\cal F} = \frac {d \; (1 + w)} {2 \; \gamma \lambda_{qm}} \;
(t - t_{ub})
\; \; \; , \; \; \;  \; \; \; 
f^2 = \frac {1} {1 + c_1 (t - t_{ub})^2}
\]
where $t = t_{ub}$ is the time when $m = \frac {\pi} {2} \;$ and
${\cal F} = 0 \;$. The scale factor $a(t)$ is given by equation
(\ref{1iso3}).

It is clear that $f \le 1$ and the scale factor $a$, and hence
the volume $V$, will not vanish. Also, the density $\rho$ will
not diverge. It is straightforward to see that as $t$ ranges
between $+ \infty$ and $- \infty$, $\; m$ ranges between $0 \;$
and $\pi \;$. When $m$ is near $0$ or $\pi \;$, it is easy to
see that the evolution is same as that given by Einstein's
equations.

%\newpage

\vspace{2ex}

\begin{center}

{\bf (ii) $\; \; \mathbf f(m) = \frac {\alpha m} {m_*} \;
\left( 2 - \frac {m} {m_*} \right) \;$}

\end{center}

\vspace{2ex}

In the case where $f(m) = \alpha z \; (2 - z) \;$ and $z = \frac
{m} {m_*} \;$, one obtains that
\[
{\cal F} = \frac {m_*} {4 \alpha^2} \; \left(
\frac {2 \; (1 - z)} {z \; (2 - z)} + ln \frac {2 - z} {z}
\right) \; = \; \frac {d \; (1 + w)} {2 \; \gamma \lambda_{qm}}
\; (t - t_0) + {\cal F}_0 \; \; ,
\]
which can also be written as 
\[
{\cal F} = \frac {m_*} {4 \alpha^2} \; \left(
\frac {2 \; (1 - z)} {z \; (2 - z)} + ln \frac {2 - z} {z}
\right) \; = \; \frac {d \; (1 + w)} {2 \; \gamma \lambda_{qm}}
\; (t - t_{ub})
\]
where $t = t_{ub}$ is the time when $m = m_*, \; z = 1$, and
${\cal F} = 0 \;$. The scale factor $a(t)$ is given by equation
(\ref{1iso3}).

We can not express $f(m)$ analytically in terms of ${\cal F}$ in
this case. The features of the evolution can, however, be read
off easily. As $m \to 0$, we have $z \to 0$ and $t \to + \infty
\;$. As $m \to m_*$, we have $z \to 1$ and $t \to t_{ub} \;$. As
$m \to 2 m_*$, we have $z \to 2$ and $t \to - \infty \;$. Thus,
as $t$ decreases from $+ \infty$ to $t_{ub}$ to $- \infty \;$,
we have that $f(m)$ increases from $0$ to $\alpha$ and then
decreases to $0$ and, hence from equation (\ref{1iso3}), that
the scale factor $a(t)$ decreases from $+ \infty \;$ to a non
zero minimum and then increases to $+ \infty \;$. The volume $V$
will not vanish and the density $\rho$ will not diverge. When
$m$ is near $0$ or $2 m_* \;$, it is easy to see that the
evolution is same as that given by Einstein's equations.

%\newpage

\vspace{2ex}

\begin{center}

{\bf (iii) $\; \; \mathbf f(m) = \alpha \; \left( 1 -
\left( 1 - \frac {m} {m_*} \right)^{2 n} \right) \;$}

\end{center}

\vspace{2ex}

In the case where $f(m) = \alpha (1 - (1 - z)^{2 n}) \;$, $\; z
= \frac {m} {m_*} \;$, and $n$ is a positive integer, we can not
obtain ${\cal F}$ explicitly for arbitrary values of $n \;$. The
features of the evolution are, however, similar to those in the
previous case and can be read off easily. As $m \to 0$, we have
$z \to 0$, $\; f \sim z $, and
\[
{\cal F} \; \sim \; - \int \frac {d z} {z^2} \; \sim \; t \; \to
\; + \infty \; \; .
\]
As $m \to 2 m_*$ from below, we have $z \to 2$ from below, $f
\sim (2 - z)$, and
\[
{\cal F} \; \sim \; - \int \frac {d z} {(2 - z)^2} \; \sim \; t
\; \to \; - \infty \; \; .
\]
As $m \to m_*$, we have $z = 1 - x \to 1$ and $t \to t_{ub}
\;$. Then $f = \alpha (1 - x^{2 n}) \;$ and 
\[
{\cal F} = \frac {m_*} {4 \alpha^2} \; (x + \cdots) \; = \;
\frac {d \; (1 + w)} {2 \; \gamma \lambda_{qm}} \; (t - t_{ub})
\; \; .
\]
Thus, as $t$ decreases from $+ \infty$ to $t_{ub}$ to $- \infty
\;$, we have that $f(m)$ increases from $0$ to $\alpha$ and then
decreases to $0$ and, hence from equation (\ref{1iso3}), that
the scale factor $a(t)$ decreases from $+ \infty \;$ to a non
zero minimum and then increases to $+ \infty \;$. The volume $V$
will not vanish and the density $\rho$ will not diverge. When
$m$ is near $0$ or $2 m_* \;$, it is easy to see that the
evolution is same as that given by Einstein's equations.

%\newpage

\vspace{2ex}

\begin{center}

{\bf (iv) $\; \; \mathbf f(m) = \alpha \; (1 - e^{- \beta m})
\;$}

\end{center}

\vspace{2ex}

Consider the case where $f(m) = \alpha \; (1 - e^{- \beta m})
\;$ and $\alpha$ and $\beta$ are positive constants. For $m \ge
0 \;$, the function starts linearly and increases monotonically
to a constant value $\alpha \;$ as $m \to \infty \;$. Unlike
other cases, in this case the function $f(m)$ approaches the
maximum asymptotically and does not decrease. It is
straightforward to perform the intgeration $\int \frac {d m}
{f^2}$ and one obtains that,
\[
{\cal F} = \frac {1} {\alpha^2 \; \beta} \; \left( \frac {1}
{1 - e^{- \beta m}} - ln \; (e^{\beta m} - 1) \right) \; = \;
\frac {d \; (1 + w)} {2 \; \gamma \lambda_{qm}} \; (t - t_0)
+ {\cal F}_0 \; \; .
\]
The scale factor $a(t)$ is given by equation (\ref{1iso3}).

We can not express $f(m)$ analytically in terms of ${\cal F}$ in
this case. The features of the evolution can, however, be read
off easily. As $m \to 0$, we have ${\cal F} \to \left( \frac {1}
{\alpha^2 \beta^2} \right) \frac {1} {m}$ and $t \to + \infty
\;$. As $m$ increases, ${\cal F}$ decreases monotonically and,
as $m \to + \infty$, we have ${\cal F} \to - \; \frac {m}
{\alpha^2}$ and $t \to - \infty \;$. Thus, as $t$ decreases from
$+ \infty$ to $- \infty \;$, we have that $f(m)$ increases
monotonically from $0$ to $\alpha$ and, hence from equation
(\ref{1iso3}), that the scale factor $a(t)$ decreases
monotonically from $+ \infty \;$ to a non zero minimum. The
volume $V$ will not vanish and the density $\rho$ will not
diverge. When $m$ is near $0$, it is easy to see that the
evolution is same as that given by Einstein's equations. Note
that, as $m \to \infty$, $f(m)$ in this case approaches the
maximum asymptotically and does not decrease from its maximum.
It is easy to see that the density $\rho$ and the scale factor
$a$ reach their finite, non zero constant values asymptotically.
Consequently, there is no bounce where the scale factor
increases again from its minimum. This phase of the evolution,
where the density and volume remain constant, is similar to the
`Hagedorn phase' in string/M theory where, as one goes back in
time, the universe's temperature $\to {\cal O}(l_s^{- 1})$ and
its density $\to {\cal O}(l_s^{- (d + 1)}) \;$ asymptotically,
$\; l_s$ being the string length scale \cite{bowick} --
\cite{k07}, \cite{k10}.

%\newpage

\vspace{4ex}

\begin{center}

{\bf 5. Conclusion}

\end{center}

\vspace{2ex}

We now summarise the present paper. We consider a $(d + 1)$
dimensional homegeneous anisotropic universe. In Einstein's
theory, it has generically a big-bang singularity in the
past. In $(3 + 1)$ dimensions, its dynamics is modified by LQC
and then it has generically a big bounce in the past, instead of
a big-bang singularity. This dynamics, modified by the quantum
effects, can be well described by effective equations of motion.

In this paper, we generalise these effective equations to $(d +
1)$ dimensions. They may then describe the modified dynamics of
a $(d + 1)$ dimensional homogeneous anisotropic universe. The
generalisation is natural and straightforward but empirical, and
involves two functions. These generalised equations may be
considered as a class of LQC -- inspired models for $(d + 1)$
dimensional early universe cosmology.

The matter Hamiltonian, in both LQC and in the models presented
here, may include various types of scalar fields and other
matter fields. But it is assumed to depend only on $p_i$ and not
on $c_i \;$. Since $c_i$ is related to $(a_i)_t \;$, this means
that matter fields couple to the metric fields but not to the
curvatures. This assumption also leads to the standard
conservation equation (\ref{t23}) irrespective of what $H_{grav}
\;$.

Special cases of the functions in the present models lead to
Einstein's equations in $(d + 1)$ dimensions and to the
effective LQC equations in $(3 + 1)$ dimensions. One can also
obtain a universe which has neither a big bang singularity nor a
big bounce but approaches asymptotically, as $t \to - \infty
\;$, a `Hagedorn like' phase where its density and volume remain
constant. In a few special cases, we also obtain explicit
solutions to the equations of motion.

\vspace{1ex}

We conclude now by mentioning a few issues for further studies.

\vspace{1ex}

{\bf (a)} In LQC, as well as in our generalisations, matter
sector remains `classical'. Quantum effects in the matter sector
should also be included.

\vspace{1ex}

{\bf (b)} In Einstein's theory, an $(n + 3 + 1)$ dimensional
universe with $n$ compact spatial directions of sufficiently
small sizes may be thought of as effectively a $(3 + 1)$
dimensional universe but with extra scalar fields appearing in
the matter sector which describe the sizes of the compact
directions. There must be an analog of this in the present
generalised equations which, however, is not clear to
us. Understanding this effective lowering of dimensions may
provide an insight into the issue {\bf (a)} mentioned above.

\vspace{1ex}

{\bf (c)} One may also obtain LQG -- inspired modifications to
Oppenheimer -- Volkoff equations which decsribe the static
spherically symmetric stars. One can then study their effects
on, for example, the maximum mass of a stable star. Similary for
Oppenheimer -- Snyder equations for stellar collapse.

\vspace{1ex}

{\bf (d)} The previous two issues may both be subsumed by finding LQG
-- inspired modifications to the $(d + 1)$ dimensional Einstein's
equations (\ref{r11}) in a covariant form. See \cite{covariant,
08shyam, helling} for a study of such modifications to the $(3 + 1)$
dimensional Einstein's equations in a covariant form, obtained from
higher curvature effective actions. Such modified covariant equations,
even if obtained only at an empirical level, can be used in a variety
of other contexts also, for example in studying the evolution of the
inhomogeneous perturbations in a universe undergoing bounce.

%\newpage 

\vspace{4ex}

{\bf Acknowledgement:} 
We thank G. Date for discussions and helpful suggetsions, and
Arnab Priya Saha for pointing out the references \cite{th1, th2,
th3}.

%\newpage

\end{document}